\documentclass[useAMS,usenatbib]{mn2e}

\usepackage{graphicx}

\title[Photometric study of IP V647 Aur]{An extensive photometric study of the recently discovered intermediate polar V647~Aur (1RXS J063631.9+353537)}

\author[V. P. Kozhevnikov]{V. P. Kozhevnikov\thanks{E-mail:valery.kozhevnikov@urfu.ru}\\
Astronomical Observatory, Ural Federal University, Lenin Av. 51, Ekaterinburg, 620083, Russia}
 
\begin{document}

\date{Accepted. Received; in original form}

\pagerange{\pageref{firstpage}--\pageref{lastpage}} \pubyear{}

\maketitle

\label{firstpage}

\begin{abstract} 
We report the results of photometry of the intermediate polar V647~Aur. Observations were obtained over 42 nights in 2012 and 2013. The total duration of the observations was 246~h. We clearly detected three oscillations with periods of $932.9123\pm0.0011$, $1008.307\,97\pm0.000\,38$ and $1096.955\pm0.004$~s, which may be the white dwarf spin period and two orbital sidebands, accordingly. The oscillation with a period of 932.9123~s has a quasi-sinusoidal pulse profile with a slightly changeable semi-amplitude from 10.9~mmag in 2012 to 12.5~mmag in 2013.  The oscillation with a period of 1008.307\,97~s has a slightly asymmetric pulse profile with a remarkable small hump on the ascending part. The semi-amplitude of this oscillation is highly changeable both in a time-scale of days (26--77~mmag) and in a time-scale of years (47~mmag in 2012 and 34~mmag in 2013). The oscillation with a period of 1096.955~s has a highly asymmetric pulse profile with a semi-amplitude of about 6~mmag. The three detected oscillations imply an orbital period of $3.46565\pm0.00006$~h. By comparing our data with the data of B.~T.~G\"{a}nsicke et al., which were obtained 8 years ago, we discovered that the spin period of the white dwarf in V647~Aur decreases with $dP/dt=(-1.36\pm0.08)\times10^{-10}$. This important result should be confirmed by future observations. Our oscillation ephemeredes and times of maxima can be useful for this confirmation. 
\end{abstract} 

\begin{keywords}
stars: individual: V647~Aur -- stars: novae, cataclysmic variables -- stars: oscillations.
\end{keywords}

\section{INTRODUCTION}

Intermediate polars (IPs) form a sub-class of cataclysmic variables (CVs), in which a magnetic white dwarf accretes material from a late type companion filling its Roche lobe. The rotation of the white dwarf is not phase-locked to the binary period of the system. Because the magnetic axis is offset from the spin axis of the white dwarf, this causes oscillations in the X-ray and optical wavelength bands. The X-ray oscillation period is usually identified as the spin period of the white dwarf. In addition to the spin and orbital periods, the reprocessing of X-rays at some part of the system that rotates with the orbital period gives rise to emission that varies with the beat period, where $1/P_{\rm beat}=1/P_{\rm spin}-1/P_{\rm orb}$. This synodic counterpart is often called the orbital sideband of the spin frequency or $\omega-\Omega$, where $\omega=1/P_{\rm spin}$ and  $\Omega=1/P_{\rm orb}$. Other orbital sidebands such as $\omega-2\Omega$ and $\omega+\Omega$ can be additionally observed \citep{warner86}. A comprehensive review of IPs is given in \citet{patterson94}.

\cite{gansicke05} reported the results from a search of CVs using a combined X-ray ({\it ROSAT}) and infrared (2MASS) target selection. They found four new IPs. One of them was denominated V647~Aur. \citeauthor{gansicke05} detected two optical oscillations with periods of $1008.3408\pm0.0019$ and $930.5829\pm0.0040$~s. In addition, from radial velocity measurements \citeauthor{gansicke05} found $P_{\rm orb}=201\pm8$~min. Recently \cite{shears11} confirmed the larger period, which turned out equal to $1008.3\pm0.5$~s.  However, until lately in the IP home page (http://asd.gsfc.nasa.gov/Koji.Mukai/iphome) this IP was reckoned among probable IPs but not among confirmed or ironclad IPs, because there was no X-ray confirmation of the period. This problem might be solved due to recent X-ray observations of V647~Aur reported by \cite{bernardini12}. They, however, detected the X-ray periods, which were incompatible with the periods found by \cite{gansicke05} and \cite{shears11}. Thus, the X-ray observations reported by \citeauthor{bernardini12} made V647~Aur a puzzling star rather than a confirmed IP.

We made a trial photometric observation of V647~Aur and discovered that it showed a short-period oscillation directly in the light curve. This oscillation was also easily detectable in the power spectrum. Therefore, V647~Aur should be no problem star. Moreover, due to large amplitude, the period of this oscillation can be measured with high precision allowing long-term tracking of period changes. To obtain the oscillation period with high precision and derive an oscillation ephemeris with a long validity, we performed extensive photometric observations of V647~Aur.  In this paper we present results of all our observations, spanning a total duration of 246~h within 42 nights.

\section{OBSERVATIONS} \label{observations}

In observations of CVs we use a multi-channel photometer that allows us to make continuous brightness measurements of two stars and the sky background. Because the angular separation between the programme and comparison stars is small, such differential photometry allows us to obtain magnitudes, which are corrected for first order atmospheric extinction and for other unfavourable atmospheric effects (unstable atmospheric transparency, light absorption by thin clouds etc.). Moreover, we use the CCD guiding system, which enables precise centring of the two stars in the diaphragms to be maintained automatically. This greatly facilitates the acquisition of long continuous light curves and improves the accuracy of brightness measurements. The design of the photometer is described in \citet{kozhevnikoviz}.

V647~Aur was observed in 2012 January--March and December and in 2013 January--March over 42 nights using the 70-cm telescope at Kourovka observatory, Ural Federal University. A journal of the observations is given in Table~\ref{journal}. Below, for brevity, the data obtained in 2012 December are ascribed to the data obtained in 2013. The programme and comparison stars were observed through 16-arcsec diaphragms, and the sky background was observed through a 30-arcsec diaphragm. The comparison star is USNO-A2.0 1200-04971757. It has $\alpha=06^h36^m41\fs48$, $\delta=+35\degr31\arcmin54\farcs86$ and $B=14.6$~mag. Data were collected at 8-s sampling intervals in white light (approximately 300--800~nm), employing a PC-based data-acquisition system.

\begin{table}
{\small 
\caption{Journal of the observations.}
\label{journal}
\begin{tabular}{@{}l c c}
\hline
\noalign{\smallskip}
Date  &  HJD start & length \\
(UT) & (-245\,0000) & (h) \\
\hline
2012 Jan 24   & 5951.316\,444 &  6.3  \\
2012 Jan 25   & 5952.134\,132 & 10.1  \\
2012 Jan 26   & 5953.164\,534 & 9.0   \\
2012 Jan 27   & 5954.120\,988 & 7.9  \\
2012 Jan 28   & 5955.102\,313 & 5.1  \\
2012 Feb 11   & 5969.120\,096 & 4.6  \\
2012 Feb 12   & 5970.118\,565 & 6.1  \\
2012 Feb 13    & 5971.118\,094 & 7.8 \\
2012 Feb 14    & 5972.118\,568 & 8.7 \\
2012 Feb 15    & 5973.119\,849 & 9.2  \\
2012 Feb 16   & 5974.118\,626 & 9.3  \\
2012 Feb 17    & 5975.123\,681 & 7.3  \\
2012 Feb 18    & 5976.132\,078 & 6.4 \\
2012 Feb 24    & 5982.135\,324 & 8.0 \\
2012 Feb 25    & 5983.137\,418 & 8.0 \\
2012 Mar 11    & 5998.199\,233 & 2.8 \\
2012 Mar 12   & 5999.156\,396 & 5.3 \\
2012 Mar 14    & 6001.234\,320 & 1.9 \\
2012 Mar 16    & 6003.171\,543 & 4.4 \\
2012 Mar 17    & 6004.165\,672 & 2.2 \\
2012 Dec 16    & 6278.321\,162 & 5.6 \\
2012 Dec 18    & 6280.252\,442 & 2.5  \\
2012 Dec 19   & 6281.302\,231 & 5.6 \\
2013 Jan 3    & 6296.126\,082 & 6.5 \\
2013 Jan 7    & 6300.300\,209 & 5.6 \\
2013 Jan 11    & 6304.095\,021 & 11.0 \\
2013 Jan 16    & 6309.215\,073 & 8.7 \\
2013 Jan 17 & 6310.268\,036 & 2.6 \\
2013 Jan 18   & 6311.289\,039 & 3.2 \\
2013 Feb 2    & 6326.229\,786 & 3.0 \\
2013 Feb 3    & 6327.256\,329 & 2.4 \\
2013 Feb 7    & 6331.153\,087 & 4.5 \\
2013 Feb 8    & 6332.117\,132 & 8.0 \\
2013 Feb 12    & 6336.119\,761& 1.8 \\
2013 Feb 16    & 6340.166\,383 & 7.3\\
2013 Mar 1      & 6353.137\,689 & 4.1\\
2013 Mar 3    & 6355.215\,228 & 4.7 \\
2013 Mar 4    & 6356.143\,816 & 3.3 \\
2013 Mar 6    & 6358.173\,664 & 6.5 \\
2013 Mar 11    & 6363.152\,900 & 6.9 \\
2013 Mar 13    & 6365.154\,998 & 6.5 \\
2013 Mar 14    & 6366.159\,642 & 5.5 \\
\hline
\end{tabular} }
\end{table}

\begin{figure}
\includegraphics[width=84mm]{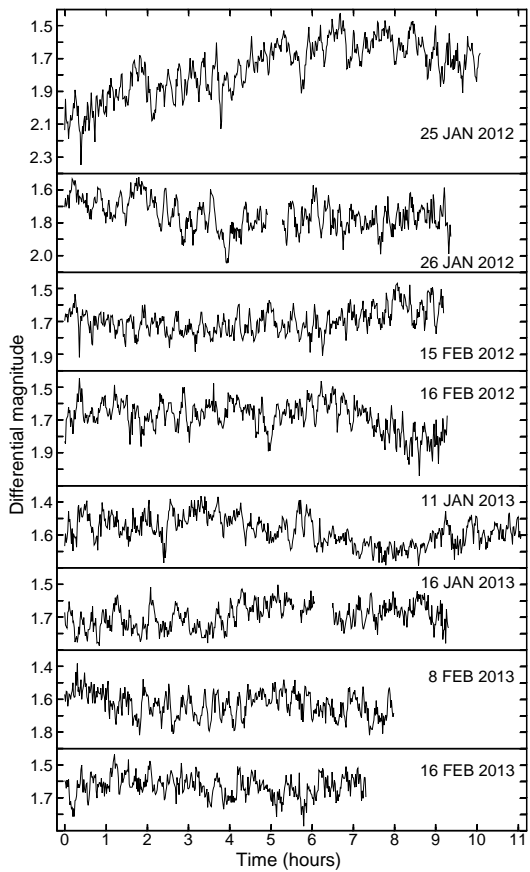}
\caption{Longest differential light curves of V647~Aur.}
\label{light}
\end{figure}

We obtained differences of magnitudes of the programme and comparison stars taking into account the differences in light sensitivity between the various channels. According to the mean counts, the photon noise (rms) of the differential light curves is equal to 0.09--0.11~mag (a time resolution of 8 s). The actual rms noise also includes atmospheric scintillations and the motion of the star images in the diaphragms. But these noise components give only insignificant additions. Fig.~\ref{light} presents the longest differential light curves of V647~Aur, with magnitudes averaged over 64-s time intervals. The white-noise component of these light curves is 0.03--0.04~mag.

\begin{figure}
\includegraphics[width=84mm]{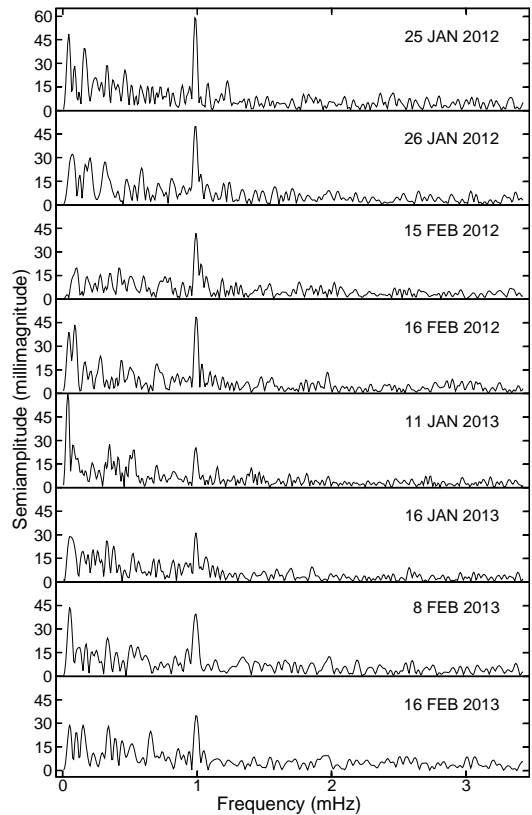}
\caption{Amplitude spectra of V647~Aur. The prominent peak visible in the amplitude spectra has a period of 1008~s.}
\label{amplitude}
\end{figure}

\begin{figure}
\includegraphics[width=84mm]{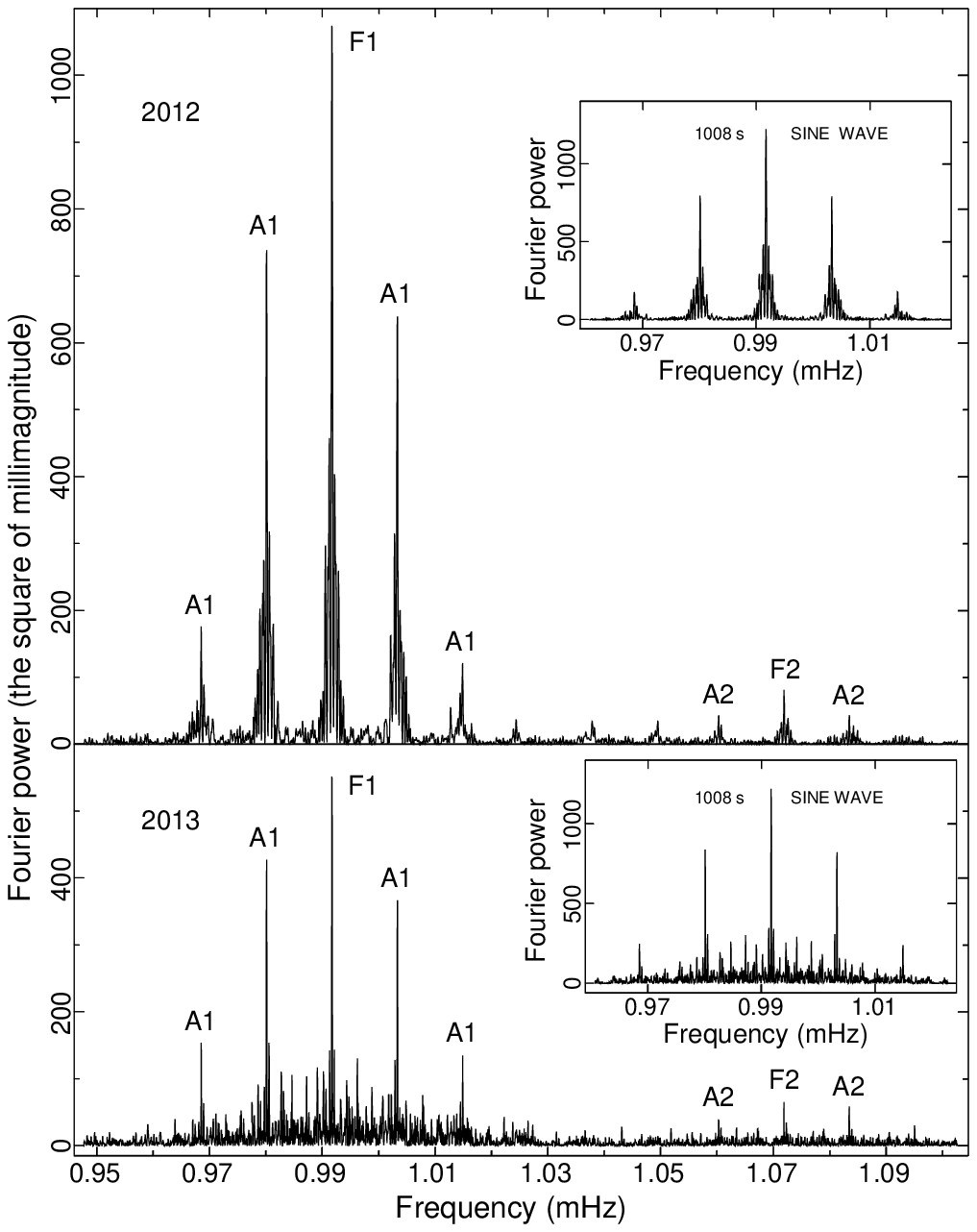}
\caption{Power spectra calculated for the data of 2012 and 2013 from V647~Aur. They reveal two coherent oscillations with the periods $P_{\rm 1}=1008.31$~s and $P_{\rm 2}= 932.9 $~s. Inserted frames show the window functions. The principal peaks are labelled with 'F1' and 'F2', and the one-day aliases are labelled with 'A1' and 'A2'.}
\label{power}
\end{figure}

\section{ANALYSIS AND RESULTS}

As seen in Fig.~\ref{light}, the light curves of V647~Aur are fairly typical of cataclysmic variables in showing rapid flickering. Moreover, a periodic oscillation is directly visible in the light curves.  Fig.~\ref{amplitude} presents the amplitude spectra calculated by a fast Fourier transform algorithm for the longest light curves of V647~Aur. Previously, low-frequency trends were removed from the light curves by subtraction of a second-order polynomial fit. The amplitude spectra reveal that the oscillation has a large semi-amplitude of about 40~mmag, which is changeable from night to night, and a period of 1008~s. 
 
One of the amplitude spectra (Fig.~\ref{amplitude}), namely 16 Feb 2012, hints at the first harmonic of the 1008-s oscillation. The average power spectrum, however, did not reveal noticeable first and second harmonics of the 1008-s oscillation. This means that the pulse profile of the 1008-s oscillation is simple and smooth. 

A distinctive feature of the periodic oscillations seen in IPs is their coherence. Analysing the data incorporated into common time series, the coherence can be demonstrated due to coincidence of the structure of the power spectrum and the window function. Fig.~\ref{power} presents the Fourier power spectra of two common time series consisting of the data obtained in 2012 and 2013. The gaps due to daylight and poor weather in these time series were filled with zeroes. Previously, low frequency trends were removed from the individual light curves by subtraction of a first or second order polynomial fit. As seen in Fig.~\ref{power}, the 1008-s oscillation reveals distinct pictures closely resembling the window functions obtained from artificial time series consisting of a sine wave and the gaps according to the observations. This proves the coherence of the oscillation over the corresponding time intervals. Due to the presence of one-day aliases, the power spectra shown in Fig.~\ref{power} allow us to detect one more coherent oscillation with much lesser amplitude.  

\citet{schwarzenberg91} showed that the $1\sigma$ confidence interval of the oscillation period is the width of the peak in the power spectrum at the $p-N^2$ level, where $p$ is the peak height and $N^2$ is the mean noise power level. We used this method to evaluate the precision of the oscillation periods. The precise maxima of the principal peaks were found by a Gaussian function fitted to the upper parts of the peaks. From the power spectra presented in Fig.~\ref{power} we found  $P_{\rm 1}=1008.3103\pm0.0061$~s and $P_{\rm 2}=932.900\pm0.016$~s in 2012 and $P_{\rm 1}=1008.3056\pm0.0042$~s and $P_{\rm 2}=932.917\pm0.009$~s in 2013. 

\begin{figure}
\includegraphics[width=84mm]{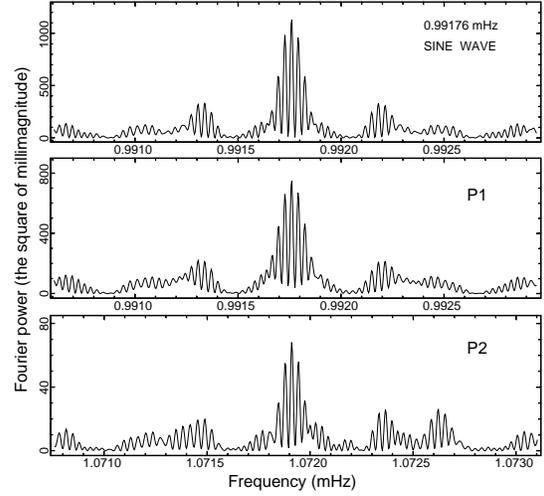}
\caption{Segments of the power spectrum calculated for all data from V647~Aur. They cover the vicinities of the periods $P_{\rm 1}$ and $P_{\rm 2}$. The upper frame shows the window function.}
\label{power2}
\end{figure}

Obviously, the highest precision of oscillation periods can be achieved from all data incorporated into common time series. The Fourier power spectrum calculated for all data from V647~Aur (Fig.~\ref{power2}) showed the distinct principal peaks well separated from different aliases for both detected oscillations. This means that our data are quite densely spaced in time, and the aliasing problem is absent. The close resemblance of the power spectrum in the vicinities of the two oscillations with the window function proves their coherence during 14 months. From this power spectrum we obtained $P_{\rm 1}= 1008.307\,97\pm0.000\,38$~s and $P_{\rm 2}=932.9127 \pm0.0011$~s.

To obtain high sampling of the Fourier power spectra, we added a considerable number of zeroes at the end of the common time series. For the power spectrum of all the data, the sampling that might allow us to measure the peak width, turned out to be extremely high, and this power spectrum consisted of $2^{25}$ points. To exclude any gross errors, which might be related with this giant number of points, we also calculated the power spectrum by a sine wave fit to the light curves folded with trial frequencies. In addition, we used the analysis-of-variance method \citep{schwarzenberg89}. By using these methods, we obtained the periods nearly identical to the periods obtained from the Fourier power spectrum. 

\begin{figure*}
\includegraphics[width=176mm]{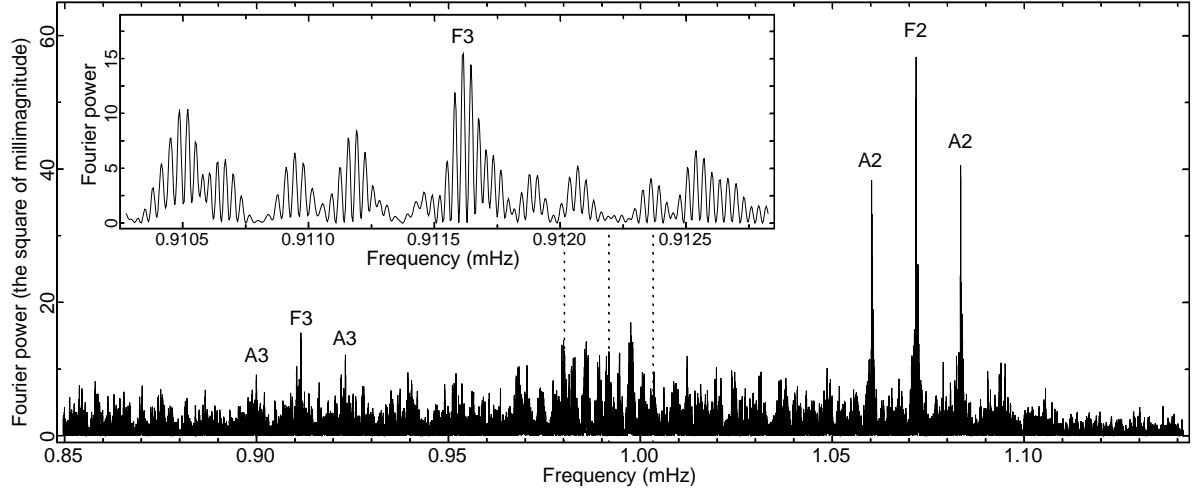}
\caption{Power spectrum of all data of V647~Aur, from which the largest oscillation was subtracted. It allows us to detect one more oscillation with the period $P_{\rm 3}$. The inserted frame shows the surrounding of the principal peak of the oscillation with $P_{\rm 3}$ on an expanded scale. The dotted lines mark the location of the principal peak of the subtracted oscillation and its one-day aliases. The one-day aliases of the oscillations with $P_{\rm 2}$ and $P_{\rm 3}$ are labelled with 'A2' and 'A3', accordingly.}
\label{prewhitened}
\end{figure*}

The power spectra (Fig.~\ref{power}) show an enlarged noise level in the vicinity of the oscillation with $P_{\rm 1}$.  This suggests the presence of additional oscillations. To find this out, we used the well-known method of subtraction of the largest oscillation from data. Indeed, the residual power spectrum (Fig.~\ref{prewhitened}) allowed us to detect one more coherent oscillation. This oscillation is obvious due to the presence of one-day aliases. This third oscillation has the period $P_{\rm 3}=1096.955\pm0.004$~s and a semi-amplitude of about 0.006~mag.

The oscillation with $P_{\rm 3}$ is quite distant in frequency from the oscillation with $P_{\rm 1}$. None the less, this oscillation is affected by the oscillation with $P_{\rm 1}$. This suggests that the oscillation with $P_{\rm 2}$ can be also affected by the oscillation with $P_{\rm 1}$. To find this out, we performed numerical experiments with artificial time series and found that, due to relatively small amplitude, the oscillation with $P_{\rm 2}$ was noticeably sensitive to the influence of the oscillation with $P_{\rm 1}$. Therefore, we remeasured the period $P_{\rm 2}$ from the pre-whitened data and found that it was equal to $932.909\pm0.016$~s in 2012, $932.916\pm0.009$~s in 2013 and $932.9123\pm0.0011$~s in all the data.

Summarized information about the periods and amplitudes of the oscillations with $P_{\rm 1}$ and $P_{\rm 2}$ obtained in different seasons is given in Tables~\ref{valuesp1} and \ref{valuesp2}. The precise semi-amplitudes and their rms errors were found from a sine wave fitted to folded light curves. In addition, in the fourth columns we give the half-width of the peaks at half maximum (HWHM), which is often used as a conservative error, and in the fifth columns we give the rms errors according to \citet{schwarzenberg91}. One can notice that the rms errors are considerably less than the conservative errors, where this difference increases with length of the observations. Obviously, growth of the precision occurs due to both the increase of the frequency resolution and the decrease of the relative noise level. Because the errors of the periods found from all the data are much lower than the other errors, we found the deviations of the other periods and expressed them in units of their rms errors. This is shown in the sixths columns of Tables~\ref{valuesp1} and \ref{valuesp2}. All the deviations are less than the rms errors. This means that the rms errors may be somewhat overestimated.

\begin{table}
{\small
\caption{The values and precisions of the period $P_{\rm 1}$.}
\label{valuesp1}
\begin{tabular}{@{}l l l l l l l }
\hline
\noalign{\smallskip}
Time   & Semi-amp. & Period & HWHM & Error & Dev. \\
span & (mmag)   & (s)       & (s)         & $\sigma$ (s) &    \\
\hline
2012 & $47\pm1$ & 1008.3103    & 0.1090   &  0.0061   & $0.4\sigma$    \\
2013 & $34\pm1$ & 1008.3056    & 0.0540   &  0.0042    & $0.9\sigma$    \\
Total  & $41\pm1$ & 1008.30797  & 0.00833  &  0.00038  & -- \\
\hline
\end{tabular} }
\end{table}

\begin{table}
{\small
\caption{The values and precisions of the period $P_{\rm 2}$.}
\label{valuesp2}
\begin{tabular}{@{}l l l l l l}
\hline
\noalign{\smallskip}
Time   & Semi-amp. & Period & HWHM & Error & Dev. \\
span & (mmag)   & (s)       & (s)         & $\sigma$ (s) &       \\
\hline
2012   & $10.9\pm0.6$  & 932.909    & 0.083    & 0.016  & $0.2\sigma$    \\
2013   & $12.5\pm0.6$  & 932.916    & 0.039    &  0.009 & $0.4\sigma$    \\
Total   & $11.7\pm0.5$  & 932.9123  & 0.0071   & 0.0011  & -- \\
\hline
\end{tabular} }
\end{table}

The individual light curves have low frequency resolution and are suitable only for the determination of the phases of the oscillation with $P_{\rm 1}$, which has much larger amplitude. However, for the light curves 3--6~h long, the relative noise can be considerable when the oscillation with $P_{\rm 1}$ diminishes its amplitude. Therefore, for the determination of the phases, we used the light curves longer than 6.9 hours (3100 points). They were folded with $P_{\rm 1}$. To precisely find the times of maxima, we used a sine wave fitted to folded light curves. Results of these phase measurements together with the oscillation semi-amplitudes found from the sine waves fitted are given in Table~\ref{maxima}. Accepting the time of maximum of the first light curve as the initial time, we derived the following tentative ephemeris:

{\small
\begin{equation}
HJD(\rm max\,1)= 245\,5952.146\,59(7)+0.011\,670\,2311(44) Å.
\end{equation} }

According to this ephemeris, we calculated the observed minus calculated (O--C) values and numbers of the oscillation cycles and presented them in Table~\ref{maxima}. The (O--C) values revealed an obvious slope. However, it seems premature to correct the ephemeris, because, as seen in Table~\ref{maxima}, the number of (O--C) values in 2013 is considerably less than in 2012, and their errors of times of maxima are larger due to less amplitude. Therefore, we decided to derive another ephemeris, for which, to obtain times of maxima, larger parts of data are used.

\begin{figure}
\includegraphics[width=84mm]{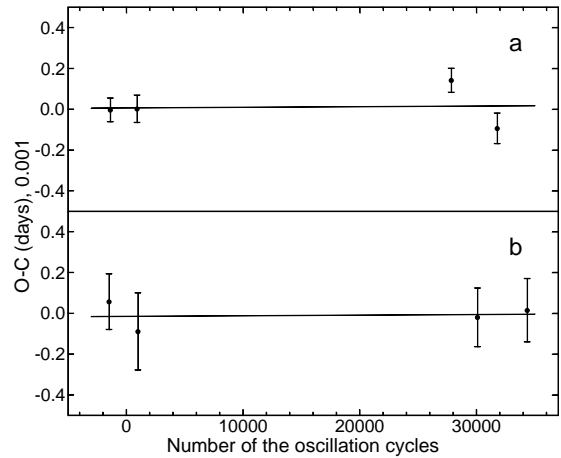}
\caption{O--C diagrams for all data from V647~Aur that are subdivided into four groups and folded with the periods $P_{\rm 1}$~(a) and $P_{\rm 2}$~(b).}
\label{ocparts}
\end{figure}

Although the individual light curves are not suitable to obtain times of maxima of the oscillation with $P_{\rm 2}$ due to its low amplitude, this seems possible by using of large parts of data. Obviously, in this case it is necessary to use pre-whitened data. By using large parts of data, we derived both ephemeris for the oscillation with $P_{\rm 1}$ and ephemeris for the oscillation with $P_{\rm 2}$. We found the initial times of maxima from the data of 2012 and utilized the data of 2013 for verification. In addition, we subdivided the data into four groups (see Tables~\ref{partsp1} and~\ref{partsp2}) and also used them for verification.  To precisely find the times of maxima and oscillation semi-amplitudes, we also used a sine wave fitted to folded light curves. Because the times of maxima were obtained from the folded light curves of large parts of data, we refer these times to the middle of corresponding observations.  Finally, for the oscillations with $P_{\rm 1}$ and $P_{\rm 2}$, we obtained the following ephemerides:
{\small
\begin{equation}
HJD(\rm max\,1)= 245\,5977.797\,720(40)+0.011\,670\,2311(44) Å.
\label{ephemeris}
\end{equation} }
{\small
\begin{equation}
HJD(\rm max\,2)= 245\,5977.799\,05(9)+0.010\,797\,596(13) Å        
\end{equation} }
According to these ephemerides, we calculated the (O--C) values and numbers of the oscillation cycles and presented them in Tables~\ref{partsp1} and~\ref{partsp2} and in Fig.~\ref{ocparts}. As seen, the (O--C) diagrams reveals only very small slopes. These slopes are much less than their rms errors. Hence, the ephemerides cannot be corrected.

\begin{table}
{\scriptsize

\caption{Verification of the ephemeris of the oscillation with $P_{\rm 1}$ by using longest individual light curves.} 
\label{maxima}
\begin{tabular}{@{}l l l l c}
\hline
\noalign{\smallskip}
{\small Date}  & {\small Semi-amp.}    & {\small HJD(max)}  & {\small N. of}    & {\small O--C}    \\
{\small (UT)}  & {\small (mmag)}   & (-245\,0000)      & {\small cycles}   &   {\small (s)}      \\
\hline
2012 Jan 25 &   $58\pm2$ &5952.146\,59(7)    &   0       &    $ -  $      \\ 
2012 Jan 26 &  $50\pm3$ & 5953.173\,63(10)  &  88      & $  5\pm10  $ \\
2012 Jan 27 &  $59\pm3$ & 5954.130\,43(8)    & 170     & $ -9\pm 9  $  \\ 
2012 Feb 13 &  $46\pm3$ & 5971.133\,77(11)  & 1627   &  $ -25\pm11  $   \\ 
2012 Feb 14 &  $76\pm2$ & 5972.126\,05(6)    & 1712   & $  2\pm 8  $        \\ 
2012 Feb 15 &  $42\pm2$  & 5973.129\,52(9)   & 1798   & $ -12\pm10  $   \\
2012 Feb 16 &  $49\pm3$  & 5974.133\,38(11)  &  1884  & $  7\pm11  $    \\
2012 Feb 17 &  $28\pm3$  & 5975.136\,58(16)  & 1970   & $   -31\pm15 $  \\ 
2012 Feb 24 & $47\pm2$   & 5982.150\,63(9)    & 2571   & $  -11\pm10 $    \\ 
2012 Feb 25 & $40\pm2$   & 5983.154\,27(9)    & 2657   &  $ -10\pm10 $    \\
2013 Jan 11 &  $26\pm3$   & 6304.109\,03(20)  & 30\,159   & $ -5\pm18  $     \\
2013 Jan 16 &  $32\pm2$   & 6309.232\,46(12)  &  30\,598  & $  12\pm12 $      \\
2013 Feb 8 &  $40\pm3$    & 6332.129\,14 (15) & 32\,560    &  $  -15\pm14 $    \\ 
2013 Feb 16 &  $35\pm2$   & 6340.181\,35(12)  & 33\,250   &  $ -36\pm12 $     \\ 
2013 Mar 11 &  $40\pm3$   & 6363.160\,17(12)  & 35\,219   & $  -25\pm12 $     \\
\hline 
\end{tabular} }
\end{table} 

\begin{table}
{\small
\caption{Verification of the ephemeris of the oscillation with $P_{\rm 1}$ by using large parts of data.} 
\label{partsp1}
\begin{tabular}{@{}l l l c}
\hline
\noalign{\smallskip}
Time         & HJD(max)          & N. of    & O--C   \\
span     & (-245\,0000)      & cycles   &   (s)    \\
\hline
2012 Jan 24--Feb 14        & 5961.902\,86(4)  & --1362   & $ -0.3\pm5 $  \\
2012 Feb 15--Mar 17        & 5988.697\,72(5)  &  934     &   $ 0.2\pm6 $  \\
2012 Dec 16--2013 Feb 3 & 6302.848\,81(4) & 27\,853  &    $ 12\pm5 $  \\ 
2013 Feb 7--Mar 14          & 6348.770\,93(6) & 31\,788  &   $ -8\pm6 $   \\ 
2013 all                             & 6322.361\,31(4) & 29\,525  &    $ 2\pm5 $    \\ 
\hline 
\end{tabular} }
\end{table} 

\begin{table}
{\small
\caption{Verification of the ephemeris of the oscillation with $P_{\rm 2}$.} 
\label{partsp2}
\begin{tabular}{@{}l l l l l}
\hline
\noalign{\smallskip}
Time         & HJD(max)          & N. of    & O--C  \\
span     & (-245\,0000)      & cycles   &   (s)  \\
\hline
2012 Jan 24--Feb 14        & 5961.905\,05(10)  & --1472  & $ 5\pm12  $  \\
2012 Feb 15--Mar 17        & 5988.693\,73(17)  &  1009   & $ -8\pm16 $  \\
2012 Dec 16--2013 Feb 3 & 6302.839\,06(11) & 30\,103   & $ -2\pm12 $  \\ 
2013 Feb 7--Mar 14          & 6348.782\,87(13)  & 34\,358   & $ 1\pm13 $  \\ 
2013 all                             & 6322.361\,13(9)    & 31\,911   &  $ 0.5\pm11 $ \\ 
\hline 
\end{tabular} }
\end{table}

\begin{figure*}.
\includegraphics[width=176mm]{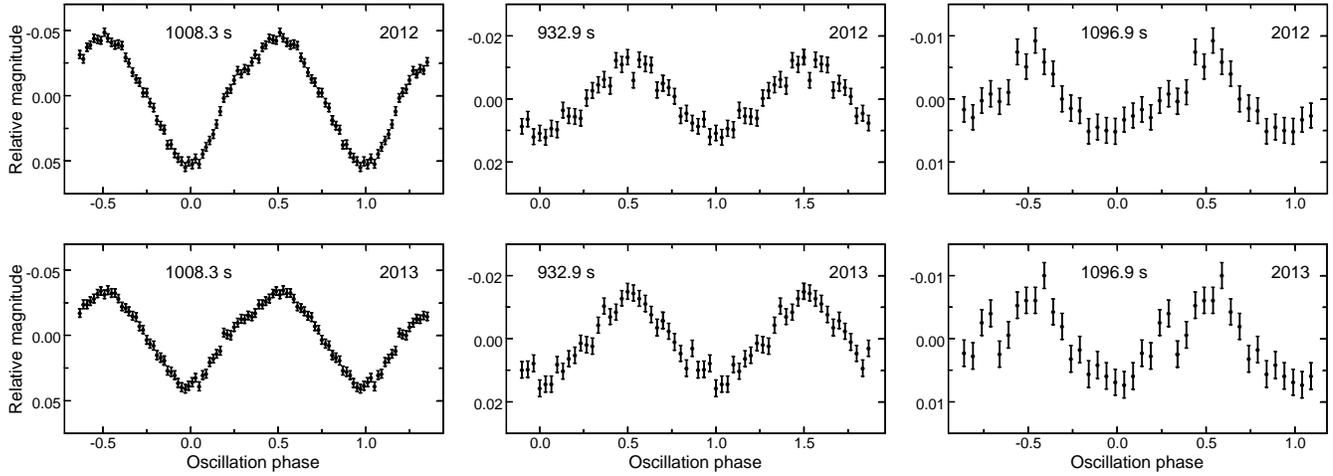}
\caption{Pulse profiles of three oscillations obtained for the data of 2012 and 2013 from V647~Aur. The oscillation with a period of 1008.307\,97~s (on the right) has a slightly asymmetric pulse profile with changeable amplitude and a small hump in phases 0.2--0.35. The oscillation with a period of 932.9123~s (in the middle) has a quasi-sinusoidal pulse profile. The oscillation with a period of 1096.955~s (on the right) reveals a highly asymmetric pulse profile with a slow rise to maximum and a rapid decline to minimum.}

\label{profiles}

\end{figure*}

The time during which the accumulated error from the period runs up to one oscillation cycle is considered as a formal validity of an ephemeris. The formal validities of the ephemerides of the oscillations with $P_{\rm 1}$ and  with $P_{\rm 2}$ are 85 and 27~yr, accordingly (a $1\sigma$ confidence level). In the second case the diminishing of the formal validity is caused by less oscillation amplitude and accordingly larger relative noise level. 

Fig.~\ref{profiles} presents the light curves of V647~Aur folded with the periods $P_{\rm 1}=1008.307\,97$~s, $P_{\rm 2}=932.9123$~s and  $P_{\rm 3}=1096.955$~s. The oscillation with $P_{\rm 1}$ has a slightly asymmetric pulse profile with a small remarkable hump in phases 0.2--0.35, where this hump is observable both in the data of 2012 and in the data of 2013. The oscillation with $P_{\rm 2}$ reveals a quasi-sinusoidal pulse profile. The oscillation with $P_{\rm 3}$ reveals a highly asymmetric pulse profile with a slow rise to maximum and a rapid decline to minimum. 

The three detected oscillation are equally spaced in frequency. Obviously, they conform to the spin period and two orbital sidebands. Then, from the most precise periods $P_{\rm 1}$ and $P_{\rm 2}$ we derive the precise orbital period, which is equal to $3.46565\pm0.00006$~h. This period coincides with the orbital period found by \cite{gansicke05} from radial velocity measurements ($3.35\pm0.13$~h). However, we could not find the orbital period in our data. This suggests that the orbital inclination of V647~Aur is less than $50^\circ$ (e.g. \citealt{ladous94}).

\section{DISCUSSION}
We performed extensive photometric observations of V647~Aur with a total duration of 246~h in 2012 and 2013 and clearly found three highly coherent oscillations with the periods $P_{\rm 1}=1008.307\,97\pm0.000\,38 $~s, $P_{\rm 2}=932.9123\pm0.0011$~s and $P_{\rm 3}=1096.955\pm0.004$~s. On average, the semi-amplitude of the oscillation with $P_{\rm 1}$ is 41~mmag, but it is highly changeable both in a time-scale of days and in a time-scale of years (see Tables~\ref{valuesp1} and \ref{maxima}). In contrast, the oscillation with $P_{\rm 2}$ reveals only small changes of its amplitude  in a time-scale of years (see Table~\ref{valuesp2}). On average, its semi-amplitude is 11.7~mmag. The semi-amplitude of the oscillation with $P_{\rm 3}$ is roughly 6~mmag and appears stable. This third oscillation was detected by us for the first time.

Two optical oscillations that certify V647~Aur as an IP were discovered by \citet{gansicke05}. We found, however, that only one of their periods is close to our 1008-s period. Another period, which is equal to $930.5829\pm0.0040$~s, is not agreed with our shortest period. The difference is 2.3~s. To realise the reason for this large difference, we simulated the observations of V647~Aur by \citeauthor{gansicke05} according to their table~1 and found that, due to unfavourable distribution of the observations in time with a large 4-month gap, the oscillation with $P_{\rm 2}$ is highly affected by the oscillation with $P_{\rm 1}$. Therefore, \citeauthor{gansicke05} could not achieve high precision of the period of this oscillation without pre-whitening. But they did not mention about such a method.

These numerical experiments demonstrated that the oscillation with $P_{\rm 2}$ cannot considerably affect the oscillation with $P_{\rm 1}$. Therefore, the larger period found by \citeauthor{gansicke05} seems reliable. It is equal to $1008.3408\pm0.0019$~s and also differs from our 1008-s period, where the difference is 0.0328~s (17$\sigma$). Because this difference is not too large, we can attribute it to the change of $P_{\rm spin}$. Dividing this difference by the middle time interval between the observations by \citeauthor{gansicke05} and our observations, which is 7.6~yr, we derive $dP/dt=(-1.36\pm0.08)\times10^{-10}$.

\citeauthor{gansicke05} gave the times of maxima for their eight light curves of V647~Aur (see their table~4). Therefore, we can check $dP/dt$ by using the following formula \citep{breger98}:
\begin{equation}
(O-C)=0.5 \, \frac{1}{p} \, \frac{dP}{dt} \, t^2.
\label{breger}
\end{equation}
From our ephemerid we obtained inconsistent values of (O--C). The average (O--C) values were equal to $+320\pm39$~s for four nights in October 2004 and $-319\pm8$~s for four nights in February--March 2005. Obviously, the (O--C) values exceeded one oscillation cycle. Indeed, according to $dP/dt$, the (O--C) value becomes equal to one oscillation cycle in 3.9~yr. Supposing that the true (O--C) value comprises three oscillation cycles plus the observed (O--C) value, two values of $dP/dt$ derived from the times of maxima in October 2004 and in February--March 2005 can be agreed. They are equal to $-1.26\times10^{-10}$ and $-1.12\times10^{-10}$, accordingly. The average difference between the $dP/dt$ found from the times of maxima and the $dP/dt$ found from the difference of periods is less than 2.1$\sigma$. This seems a good concord and proves reality of the period change.

\citeauthor{gansicke05} could not decide which of the two periods is $P_{\rm spin}$. The recent X-ray observations of V647~Aur presented by \cite{bernardini12} can help to solve this question. They found three X-ray periods. Two of them, which are equal to $920\pm1$ and $952\pm4$~s were attributed to $P_{\rm spin}$ and $\omega-\Omega$, accordingly. These periods, however, are incompatible with the optical periods. Moreover, the orbital period being derived from the two X-ray periods, which is equal to $7.6\pm0.9$~hr, is also incompatible with the orbital period found by \citeauthor{gansicke05}. Obviously, this total incompatibility signifies an error.

\begin{figure}
\includegraphics[width=84mm]{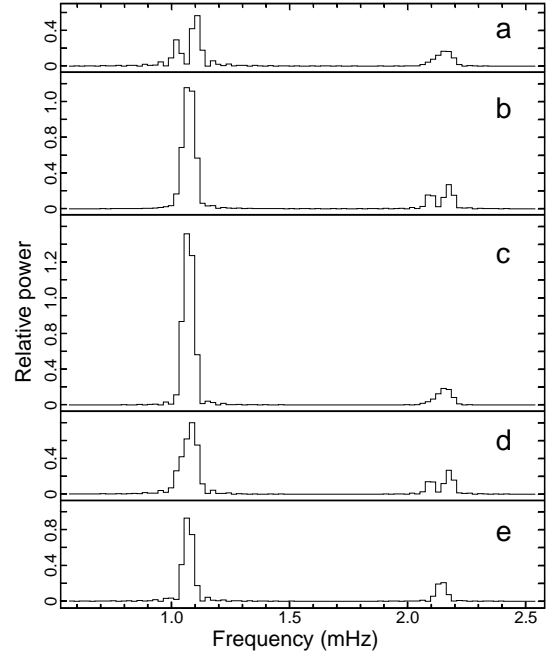}
\caption{Power spectra which characterise interaction of two oscillations in short time series. The initial phase differences are 0.2 (a), 0.5 (b), 0.7 (c) and 0.0 (d). Fig.~\ref{bernardini}e shows the power spectrum of a short time series consisting of a single sine wave with a 933-s period, to which a high-frequency harmonic was added.}
\label{bernardini}	
\end{figure}

To realise the reason for this incompatibility, we simulated the X-ray observations of V647~Aur. We constructed artificial time series consisting of two sine waves with the periods and amplitudes according to table~2 in \citeauthor{bernardini12} and with changeable initial phase differences. In fact, these phase differences are the phases of the beat period between the two sine waves. A high-frequency harmonic with half-amplitude was added to each of these sine waves. The length of these time series was 4.75~h, which is the net exposure times minus the solar flare removed exposure according to table~1 in \citeauthor{bernardini12} Fig.~\ref{bernardini} presents  typical Fourier power spectra calculated with frequency bins of 0.0222~mHz (which are seen in the X-ray power spectrum of  V647~Aur). The effects of interaction of two oscillations in short time series turned out quite diverse. Depending on the included phases of the beat period, two oscillations can be resolved (Fig.~\ref{bernardini}a) or unresolved (Figs.~\ref{bernardini}b, c). Moreover, the peak can be asymmetric and thus hinting at two oscillations (Fig.~\ref{bernardini}d). The probability of this particular case is rather low and equal roughly to 1/5.

\citeauthor{bernardini12} suspected two oscillations due to asymmetry of the peak in the X-ray power spectrum (see their fig.~3). Indeed, two oscillations with the suggested periods and amplitudes can produce a single asymmetric peak with a rather low probability. But this peak must be more wide than seen in the X-ray power spectrum if the length of the observations is 4.75~h (Fig.~\ref{bernardini}d). Hence, the asymmetry of the peak in the X-ray power spectrum must be attributed to noise rather than the two oscillations. Moreover, the high-frequency harmonic visible in the X-ray power spectrum conforms also to a single oscillation because, otherwise, the corresponding peak must be either noticeably wider or resolved into two components. Obviously, \citeauthor{bernardini12} also believe that this peak conforms to a single oscillation because, in their fig.~3, it is labelled with '2$\omega$'. Directly from the X-ray power spectrum of V647~Aur presented in fig.~3 in \citeauthor{bernardini12} we found the first harmonic is equal to $2.141\pm0.011$~mHz. Here the error is half of the frequency bin. Then $P_{\rm spin}$ is equal to $934\pm5$~s and conforms pretty well to $P_{\rm 2}$. Note that latter conclusion remain valid even if the X-ray observations are longer than 4.75~h. 

\citeauthor{bernardini12} reported detection of the third X-ray oscillation in V647~Aur, insisting that it is $\omega-2\Omega$. However, they themselves recognize that their estimated orbital period (7.6~h) is inconsistent with $\omega-2\Omega$ (see page 8 in \citeauthor{bernardini12}). We found that the period corresponding to the left blue dotted line denoting $\omega-2\Omega$ in fig.~3 in \citeauthor{bernardini12} is equal to $1014\pm11$~s. This period also conforms pretty well to $P_{\rm 1}$. Thus, the careful inspection of the X-ray power spectrum convinces it shows only two periods rather than three periods, where these periods coincide with the optical periods. Then we conclude that $P_{\rm 2}$ is $P_{\rm spin}$ because it coincides with the shortest X-ray period having the largest X-ray amplitude. Consequently, the oscillation with $P_{\rm 1}$ is $\omega-\Omega$, and the oscillation with $P_{\rm 3}$ is $\omega-2\Omega$.

Two-pole disc-fed accretion is believed to be the normal mode of behaviour in IPs. Depending on the sizes of the accretion curtains both single-peaked and double-peaked spin pulse profiles can be produced. In IPs with strong magnetic fields two accreting poles can act in phase so that single-picked, roughly sinusoidal pulse profiles can be produced, whereas in IPs with weak magnetic fields two accreting poles can act in anti-phase and produce double-peaked spin pulse profiles \citep{norton99}. It is considered that the rapidly spinning IPs with $P_{\rm spin} <700$~s have weak magnetic fields and therefore usually produce double-peaked pulse profiles (\citeauthor{norton99}). According to $P_{\rm spin}$, V647~Aur must have a strong magnetic field and, therefore, it must show a quasi-sinusoidal spin pulse profile. This conception agrees with the quasi-sinusoidal pulse profile of the oscillation with $P_{\rm 2}$.

In the case of disc-fed accretion the optical orbital sideband arises due to reprocessing of X-rays by structures rotating in the reference frame connected with the orbital motion. In particular, reprocessing can occur in asymmetric parts of the disc. V647~Aur shows highly changeable amplitude of the sideband oscillation. Large changes of the sideband amplitude in a time-scale of days are especially surprising (see Table~\ref{maxima}). These changes are not accompanied by variations of the star brightness (Fig.~\ref{light}) and hence by changes of the structure of the disc. Moreover, these changes are not accompanied by changes of the amplitude of the oscillation with $P_{\rm spin}$. Therefore, the sideband oscillation in V647~Aur is difficult to account for through reprocessing. 

Another way to produce the orbital sideband is alternation of the accretion flow between two poles of the white dwarf with the sideband frequency. This process occurs in cases of stream-fed and disc-overflow accretion and is the only way to produce the strong orbital sideband in X-rays \citep{wynn92}. The change of the contribution between the disc-overflow accretion and disc-fed accretion was observed in FO~Aqr \citep{evans04} and MU~Cam \citep{staude08}. Although this change in MU~Cam was accompanied by a large change in optical brightness, such changes seem possible without brightness changes, because the disc-overflow accretion can amount to only a small part of the full accretion (e.g. \citealt*{mukai94}). The amplitude of the sideband modulation of V647~Aur in X-rays is relatively small and does not indicate inevitably disc-overflow accretion. None the less, we can account for the optical orbital sideband and changes of its amplitude in V647~Aur by disc-overflow accretion, because the contribution of disc-overflow accretion can probably change without brightness changes.
 
\citet*{norton04} used a model of magnetic accretion to investigate the spin equilibria of magnetic CVs. This allowed them to infer approximate values for the magnetic moments of most known IPs. Many authors used their fig.~2 to evaluate the magnetic moments of newly discovered IPs (e.g. \citealt{rodriguez05, katajainen10}). We also found the place of V647~Aur in fig.~2 in \citeauthor{norton04} and the corresponding magnetic moment, which is equal to $1\times10^{33}$\,G\,cm$^3$. By considering properties of its nearest near-by IPs, which are V2004~Oph,  BG~CMi, FO~Aqr and MU~Cam, we, however, discovered  that these IPs have different magnetic fields and show different accretion modes. Indeed, BG~CMi and V2004~Oph represent rare cases, showing circular polarization and indicating thus strong magnetic field \citep{katajainen07, katajainen10}, whereas FO~Aqr represent the counterexample with probably less strong magnetic field because of non-detection of circular polarisation \citep{stockman92}. None the less, FO~Aqr possesses disc-overflow accretion similar to observed in MU~Cam and BG~CMi \citep{norton92}.  However, unlike the three IPs, V2004~Oph exhibit pure stream-fed accretion due to more strong magnetic field \citep{hellier02}. The reason for this discrepancy may consist in substantial deviation from spin equilibrium. Indeed, FO~Aqr is the only IP, which clearly demonstrates the spin equilibrium by alternating spin-up and spin-down \citep{kruszewski98, patterson98, williams03}. Other sparse IPs, in which spin period changes were observed, show either continuous spin-up or continuous spin-down \citep{kruszewski98}.

We found that V647~Aur reveals decreasing of $P_{\rm spin}$ with $dP/dt=(-1.36\pm0.08)\times10^{-10}$. This change is close to the largest period changes detected in other IPs (see table~1 in \citealt{warner96}) and also close to the period change lately observed in FO Aqr ($dP/dt=-2.0\times10^{-10}$, \citealt{williams03}). Being a new result, the spin period change in V647~Aur should be confirmed by future observations. This can be made by using our ephemerides and by using our times of maxima. We note that the large $dP/dt$ observed in  V647~Aur facilitates possible detection of alternating spin-up and spin-down.

\section{CONCLUSIONS}

We performed extensive photometric observations of V647~Aur over 42 nights in 2012 and 2013. The total duration of the observations was 246~h. 
\begin{enumerate}
\item The analysis of these data allowed us to clearly detect three coherent oscillations with the periods $P_{\rm 1}=1008.307\,97\pm0.000\,38 $~s, $P_{\rm 2}=932.9123\pm0.0011  $~s and $P_{\rm 3}=1096.955\pm0.004$~s. The shortest period is the spin period of the white dwarf. Two other periods correspond to its two negative orbital sidebands.
\item These three oscillations imply that the orbital period of the system is equal to $3.46565\pm0.00006$~h. 
\item The oscillation with $P_{\rm 1}$ has a slightly asymmetric pulse profile with a small hump in the ascending part.   The oscillation with $P_{\rm 2}$ has a quasi-sinusoidal pulse profile. The oscillation with $P_{\rm 3}$ reveals a highly asymmetric pulse profile. 
\item The high precision of two periods allowed us to obtain the oscillation ephemeredes with formal validities of 85~yr for the oscillation with $P_{\rm 1}$ and of 27~yr for the oscillation with $P_{\rm 2}$. 
\item By comparing our data with the data obtained by \cite{gansicke05} 8 years ago, we discovered that V647~Aur exhibits the period decrease with $dP/dt=(-1.36\pm0.08)\times10^{-10}$. 
\end{enumerate}

\section*{Acknowledgments}

This research has made use of the SIMBAD database, operated at CDS, Strasbourg, France. This research also made use of the NASA Astrophysics Data System (ADS).

\end{document}